# DFT and Monte Carlo simulations of the equiatomic quaternary Heusler Alloy CoFeCrP

**S. IDRISSI[1,*], S. ZITI[2], H. LABRIM[3], L. BAHMAD[1,*] and A. BENYOUSSEF[4]**

[1] Laboratoire de la Matière Condensée et des Sciences Interdisciplinaires (LaMCScI), Mohammed V University of Rabat, Faculty of Sciences, B.P. 1014 Rabat, Morocco.

[2] Intelligent Artificial and Security of Systems, Mohammed V University of Rabat, Faculty of Sciences, B.P. 1014 Rabat, Morocco.

[3] USM/DERS/Centre National de l'Energie, des Sciences et des Techniques Nucléaires (CNESTEN), Rabat, Morocco.

[4] Hassan II Academy of Science and Technology, Av. Mohammed VI, Rabat, Morocco.

**Abstract:**

In this work, we study the equiatomic quaternary Heusler Alloy CoFeCrP using two methods: DFT and Monte Carlo simulations. The DFT method allowed us to illustrate the structural, electronic and magnetic properties of this alloy. The ground state phase diagrams have been presented to show the stable configurations in different physical parameter planes. On the other hand, the Monte Carlo simulations, performed under the Metropolis algorithm, permitted to deduce the critical the behavior of the equiatomic quaternary Heusler alloy CoFeCrP.

The structural properties results show that the phase of type I, of this alloy is the most stable configuration. In addition, the band structures, and density of states calculations results show that this compound exhibits a half-metallic character with a 100 % of spin polarization (SP) at the Fermi-level. The total magnetic moment of the Heusler compound is found to be 4.00 $\mu_B$. Moreover, it is found that the Slater-Pauling is well described for this alloy.

Our results show that this material is a potential candidate for the spintronic applications. This is due to its half-metallicity, its high spin moments, its complete SP polarization and its high Curie temperature.

*) Corresponding authors: samiraidrissi2013@gmail.com (S. I.); bahmad@fsr.ac.ma (L. B.)

# I. Introduction

The Heusler alloys have inspiring the researches and have pulled in extraordinary consideration by the different studies. Because these alloys have a lot of applications, for example, magneto-electronics and spintronic devices [1, 2]. Numerous Heusler alloys have an exceptional property, it is: demi-metallicity [3]. Much investigation has been executed to explore the Heusler alloys [4-6].

The quaternary Heusler compound with formula XX'YZ possess a LiMgPbSb type [7, 8] or sort Y (L21) and the space group F-43m [9-11]. Where X, X' and Y are progress transition metals and Z is a primary sp-group.

Some the co-based quaternary Heusler alloys have been expected to be half-metallic behavior, for example, CoFeMnSi [12]. Vajiheh Alijani et *al*, have examined experimentally and theoretically the equiatomic Heusler alloys: CoFeMnZ (Z=Al, Ga, Si and Ge) [13], CoFeMnZ (Z=Si, As and Sb) [14], CoRuTiZ (Z = Si, Ge and Sn) [15]. In addition, other quaternary Heusler alloys has studied by researches, such as: TiZrCoIn [16], NiCoCrGa [17], CoFeTiZ and CoFeVZ (Z = Al, Ga, Si, Ge, As and Sb) [18], ZrVTiZ (Z = Al, Ga) [19], ZrCoTiZ (Z = Si, Ge, Ga and Al) [20], ZrMnVZ (Z = Si, Ge) [21], CoMnCrAl [22], CoYCrZ (Z = Si and Ge) [23], ZrFeVZ (Z = Al, Ga, In) [24], YCoCrZ (Z = Si, Ge, Ga) [25], CrVYZ (Z = Si, Ge, Sn) and CrVScZ (Z = Si, Ge, Sn) [26] and MCoVZ (M = Lu, Y; Z = Si, Ge) [27].

The purpose of this study is to study the physical properties of the Co-based equiatomic quaternary Heusler alloy CoFeCrP. In the first step, we performed the ab-initio method using the Quantum Espresso package. We investigate the structural, the electronic properties to study the density of states, band structure of this alloy for three configurations type I, type II and type III, we present also the phase diagrams of this alloy for null temperature (T = 0). On the other hand, we have performed the Monte Carlo simulations to examine the magnetic properties and the critical behavior of this system at non null temperature. In some of our recent works, we have applied the Monte Carlo method to study the magnetic behavior of different spintronic materials, see for example Refs. [28-34].

This paper is organized as follows: in section II, we give a detailed description of the crystal structure and computational method of the ab-initio calculations in the bases of the generalized gradient approximation GGA. In part III, we illustrate the Monte Carlo method and discus the obtained results after elaborating the ground state phase diagrams for different stable configurations in different planes of different physical parameters. Part IV is dedicated to conclusions.

## II. Computational details of ab-initio method of the equiatomic quaternary Heusler alloy CoFeCrP

The equiatomic quaternary Heusler alloy CoFeCrP crystallizes in the cubic structure for the three possible geometries illustrated in Fig. 1. The geometry of the studied system belongs to the space group *F43m*, see Fig. 1. The three possible different types of atom placement in the quaternary Heusler compound are presented in table 1. The position of the atoms of the different types of this compound are drawn using the VESTA package [35], and depicted from Ref. [36].

|            | 4a (0,0,0) | 4c (1/4, 1/4,1/4) | 4b (1/2, 1/2, 1/2) | 4d (3/4, 3/4, 3/4) |
|------------|------------|-------------------|--------------------|--------------------|
| **Type (I)**   | P          | Fe                | Cr                 | Co                 |
| **Type (II)**  | P          | Cr                | Fe                 | Co                 |
| **Type (III)** | Fe         | P                 | Cr                 | Co                 |

*Table 1: The three possible different types of the atoms of the equiatomic quaternary Heusler alloy CoFeCrP [36].*

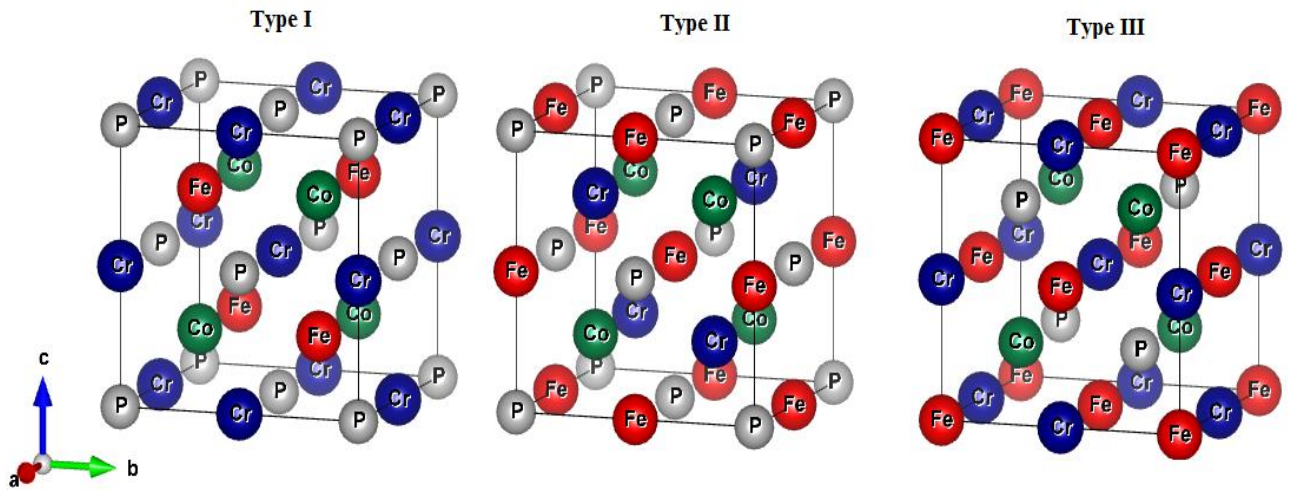

*Fig.1: The three possible geometries of the equiatomic quaternary Heusler alloy CoFeCrP.*

The structural, magnetic and electronic properties are elaborated with first-principles calculations [37, 38] based on the density functional theory (DFT) [37, 39].

We have used the spin polarized density functional theory with the pseudo-potential Plane Wave method (PP-PW) in the bases of the Quantum Espresso code [40]. The electronic exchange-correlation was treated using the generalized gradient approximation (GGA) [41] corrected by exchange and correlation functional PBE as proposed by Perdew-Burke-Ernzerhof [42]. Different recently our works have been used DFT and other method [43-46].

For the Brillouin zone integration, we take the k-point equivalent to 9x9x9. Among the self-consistency cycles, the cutoff energy, which is defined as the partition of valence and center states, was picked as 340 eV.

### a. Structural results of the equiatomic quaternary Heusler compound CoFeCrP

To predict the structural properties of this compound, we should find its optimal structure.

The minimal energy configuration of the equiatomic quaternary Heusler alloy CoFeCrP $E_f^{CoFeCrP}$ is defined as the variation of energy when this alloy is obtained from its initial elements. For the studied Heusler compound, this energy can be estimated as follow [47]:

$$E_f^{CoFeCrP} = E_{tot}^{CoFeCrP} - (E_{Co}^{Bulk} + E_{Fe}^{Bulk} + E_{Cr}^{Bulk} + E_{P}^{Bulk}) \quad (1)$$

$E_f^{CoFeCrP}$: Represents the calculated total energy at the equilibrium.

$E_{tot}^{CoFeCrP}$ : represents the total energy of the studied system.

$E_{Co}^{Bulk}$, $E_{Fe}^{Bulk}$, $E_{Cr}^{Bulk}$ and $E_{P}^{Bulk}$ : are the energies per atom of the individual pure elements: Co, Fe, Cr and P, in their specific bulk state, respectively.

The total energy as a function of lattice constant of the three types I, II and III for the equiatomic quaternary Heusler alloy CoFeCrP is clearly visible in Fig.2. From this figure, we establish that the type I structure has the lowest total energy indicating that this is the preferred crystal structure for CoFeCrP. This is in good agreement with different earlier studies of the Heusler compound, see Ref [36]. Other studies found that the type I is the more stable of the different quaternary Heusler compound structures, see Refs. [48-50].

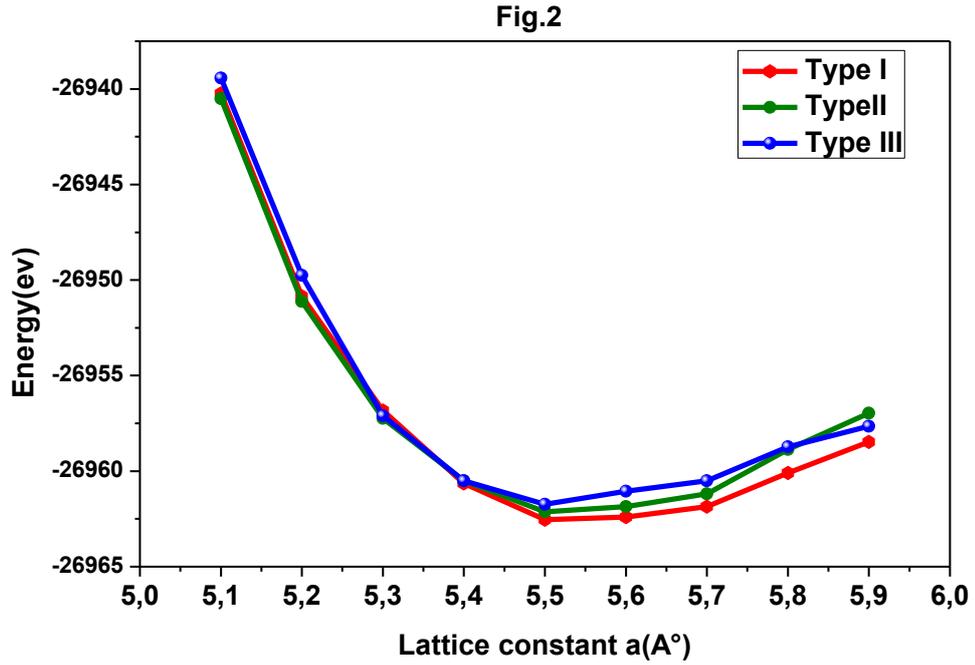

*Fig.2: The calculated total energy as a function of the lattice constant of the quaternary Heusler compound CoFeCrP for the three different structures.*

| Heusler Alloy | Types | Lattice parameter a (Å) | Optimized energy $E_o$ (eV) |
|---|---|---|---|
| CoFeCrP | **Type I** | 5.4994 | -20962.4413 |
| | **Type II** | 5.5016 | -20962. 0769 |
| | **Type III** | 5.5016 | -20961.7125 |

*Table 2: Optimized lattice parameter and energy of the equiatomic quaternary Heusler alloy CoFeCrP for the all configurations: type I, type II and type III.*

**b. Electronic and magnetic properties of the equiatomic quaternary Heusler CoFeCrP**

In this section, we study the electronic and magnetic properties for the stable configuration type I of the equiatomic quaternary Heusler alloy CoFeCrP. The density of states (DOS) of each Co, Fe, Cr and P atoms of the quaternary compound split into two spin channels is displayed in Figs. (3a) and (3b). In fact, Fig.3.a represents the calculated total DOS and partial DOSs for the configuration of the type I .

Such figure summarizes also the energy gap in the minority spin states around the Fermi level. Also, Fig.3.b confirms that this compound has a half-metallic ferromagnetic behavior.

On the other hand, we present the band structures in the Figs. 4 (a) and (b) for the two spin channels: spin-up and spin-down, respectively. The band structure of majority-spins displays the metallic behavior of the Heusler compound CoFeCrP. While, the band structure of the minority-spins allows to predict a semiconducting behavior, with a direct gap, as the valance band G presents a maximum corresponding to the lower limit of the conduction band (BC). The band structures confirm the half-metallic behavior of this Heusler alloy CoFeCrP. The energy gap of the equiatomic quaternary Heusler alloy CoFeCrP is approximately equal to $\approx 1.0$ eV. The results of the energy gap of the equiatomic quaternary Heusler alloy CoFeCrP are summarized in Table 3 for the equilibrium lattice constant $a_0 = 5.50$ A°. This table shows that our results are closer to those to Ref. [36].

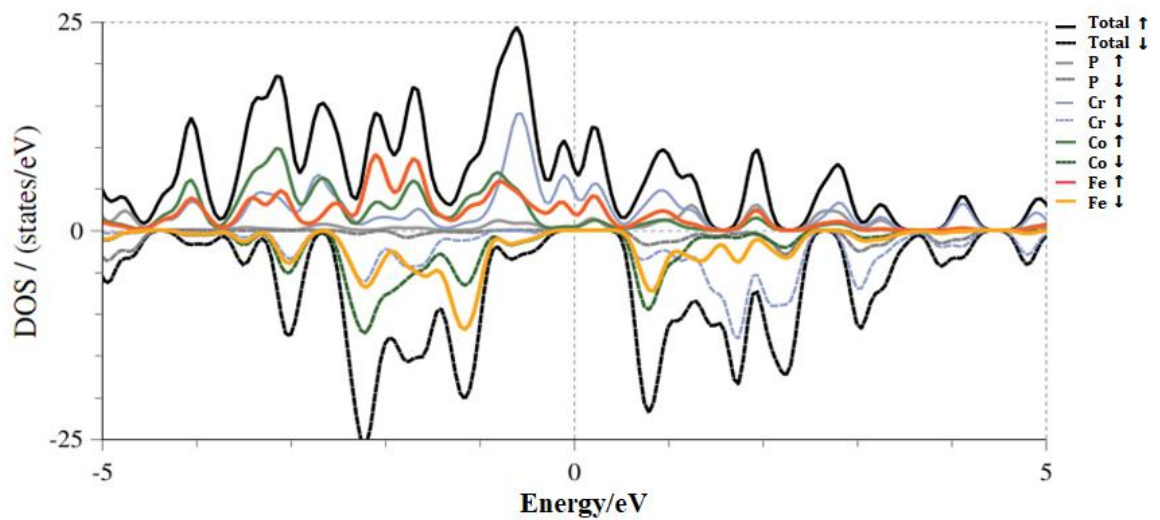

(a): Total and partial density of states of CoFeCrP alloy

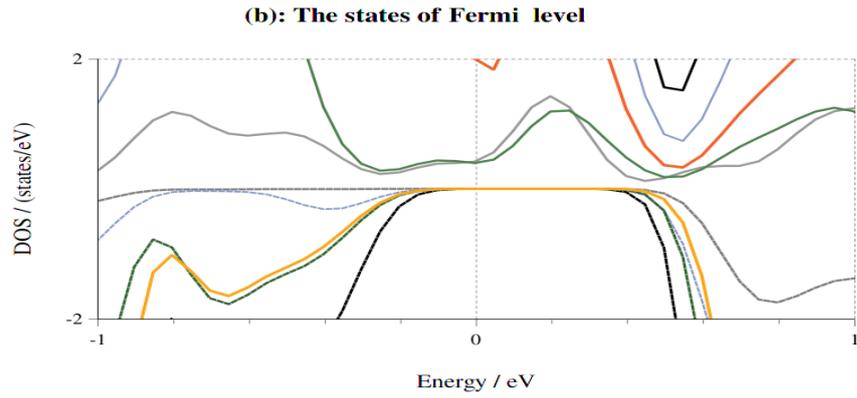

*Fig.3: (a) Calculated total and partial DOSs of the Heusler compound CoFeCrP for type I atomic arrangement fashion. (b) The states at the Fermi level.*

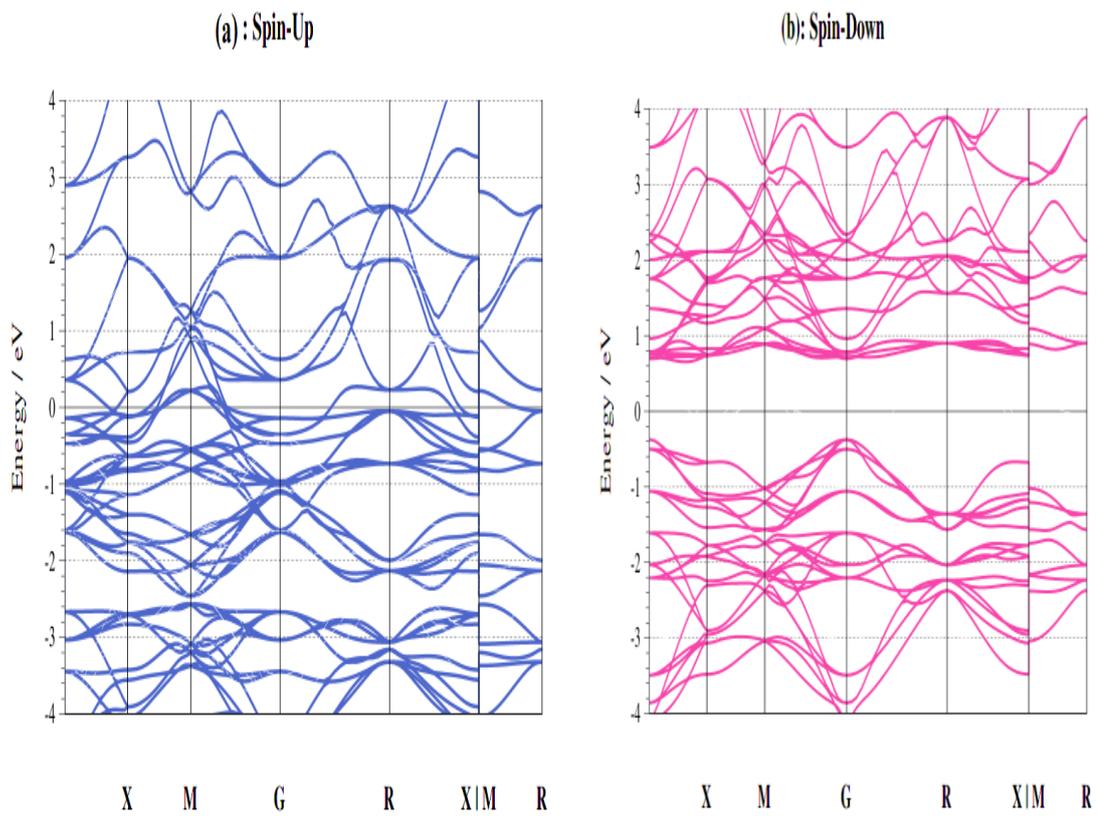

*Fig.4: Band structures of the EQH alloy CoFeCrP for the type I configuration.*

| CoFeCrP Compound | $a_0$(Å) | $E_{tot}$ (eV) | $E_g$(eV) | comportement |
|---|---|---|---|---|
| Ref.[36] | 5.59 | -110459.08 | 1.00 | Half-metallic |
| Present study | 5.50 | -107849.2 | 1.00 | Half-metallic |

*Table 3: The results of the lattice optimization of the equiatomic quaternary Heusler alloy CoFeCrP, $a_0$ is the equilibrium lattice constant, $E_{tot}$ is the total energy, $E_g$ represents the band gap.*

**C. Slater-Pauling rule and the spin polarization (SP) of the equiatomic Heusler alloy CoFeCrP**

The Slater-Pauling rule can be defined by different ways as $M_t = N_v - 18$ or $M_t = N_v - 24$. Such explanations are based on the difference of the origin of various band gaps in equiatomic quaternary Heusler alloy (EQHA). The detailed description on this alloy can be obtained in a study reported by Ozdogan *et al.* [51]. As, there is different generalized electron filling rule exists for different band gap origins [52-54].

In this study, we have calculated the $M_t$ according to the rule: $m_t = N_v - 24$.

$m_t$ : Total magnetic moment.

$N_v$ : Total number of valance electrons.

From table 4, it may be noted that the total magnetic moment of the studied alloy, exhibits partial and total magnetic moments. The valence electron configurations of the transition elements are: Co ($3d^7 4s^2$), Fe ($3d^6 4s^2$), Cr ($3d^5 4s^1$) and P ($3s^2 3p^3$). This, leads to $N_v$=28 valence electrons in the Heusler CoFeCrP alloy. As it is appeared in table 2, the total magnetic moment of this alloy is equal to 4.00 $\mu_B$. The Slater Pauling rule is well respected for this material.

The Spin Polarization can be defined as a method that can differentiate between the majority and minority states near the Fermi level. The SP of the Heusler alloys can be obtained as follows [52-54]:

$$SP = \frac{DOS(E_F(\uparrow)) - DOS(E_F(\downarrow))}{DOS(E_F(\uparrow)) + DOS(E_F(\downarrow))} \quad (2)$$

The symbols ↑ and ↓ are used for majority and minority states, respectively. While $DOS(E_F)$ represents the densities of states at the Fermi level.

The SP of the normal metal is zero (P = 0) because metals have no band gaps.

As, the Heusler alloys are completely Half metallic (HM), they have 100% SP. The results of the studied Heusler are presented in table 5. This table shows 100% SP confirming that electrons conductions can participate in just one spin channel. Due to the existence of complete SP in the equiatomic quaternary Heusler alloy CoFeCrP, it is predicted that this material may be used in the fabrication of spintronic devices.

| CoFeCrP | $m_p$ ($\mu_B$) | $m_{Co}$ ($\mu_B$) | $m_{Fe}$ ($\mu_B$) | $m_{Cr}$ ($\mu_B$) | $m_{tot}$ ($\mu_B$) |
|---|---|---|---|---|---|
| Ref. [36] | 0.014 | 1.036 | 0.904 | 1.974 | 4.00 |
| Present study | 0.08 | 1.03 | 0.87 | 2.02 | 4.00 |

*Table 4: The partial and total magnetic moments of the compound Heusler CoFeCrP.*

| Alloy | $m_{tot}$($\mu_B$) | Number of valance electrons | Slater Pauling rule magnetic moment | DOS ($E_F(\uparrow)$) | DOS ($E_F(\downarrow)$) | Spin polarization |
|---|---|---|---|---|---|---|
| CoFeCrP | 4.00 | 24 | 4.00 | 7.25 | 0 | 100 % |

*Table 5: Number of valance electrons, Slater Pauling rule magnetic moment and polarization of the equiatomic quaternary Heusler alloy CoFeCrP for the configuration structure type I.*

### III. Monte Carlo results of the equiatomic quaternary Heusler alloy CoFeCrP
#### a. Study of the ground state phases

In this section, we discuss and study the ground states phase diagrams of the Heusler compound CoFeCrP. For the disappearing temperature (T=0 K). We claim at the rendering of the effect of physical parameters on the magnetizations and the susceptibilities of this system. This calculation is established on the Hamiltonian given in Eq. (3), we utilize free limit conditions on the lattice. The obtained results in this work are given for the particular super-cell estimate 5x5x5. The different components and portions of the compound CoFeCrP are modeled by the magnetic elements: Co, Fe and Cr, and have the spin values S=2, σ=2 and Q=3/2, respectively.

$$\mathcal{H} = -J_{Co-Co} \sum_{<i,j>} S_i S_j - J_{Fe-Fe} \sum_{<i,j>} \sigma_i \sigma_j - J_{Cr-Cr} \sum_{<i,j>} Q_i Q_j -$$
$$J_{Co-Fe} \sum_{<i,j>} S_i \sigma_j - J_{Co-Cr} \sum_{<i,j>} S_i Q_j - J_{Fe-Cr} \sum_{<i,j>} Q_i \sigma_j - D_S \sum_i (S_i^2) -$$
$$D_\sigma \sum_i (\sigma_i^2) - D_Q \sum_i (Q_i^2) - H \sum_i (S_i + \sigma_i + Q_i)$$

(3)

$J_{Co-Co}$: The exchange interaction between Co-Co atoms.

$J_{Fe-Fe}$: The exchange interaction between Fe-Fe atoms.

$J_{Cr-Cr}$: The exchange interaction between Cr-Cr atoms.

$J_{Co-Fe}$: The exchange interaction between Co-Fe atoms.

$J_{Co-Cr}$: The exchange interaction between Co-Cr atoms.

$J_{Fe-Cr}$: The exchange interaction between Fe-Cr atoms.

$D_S$: is the crystal field of Co ions. $D_\sigma$: is the crystal field of Fe ions. $D_Q$: is the crystal field of Cr ions. For simplicity, we will assume that: $D = D_Q = D_S = D_\sigma$.

| | | |
|---|---|---|
| $(a_1) = (-2.0, -2.0, -1.5)$ | $(a_2) = (-1.0, -1.0, -1.5)$ | $(a_3) = (-1.0, -1.0, -0.5)$ |
| $(a_4) = (0.0, 0.0, -1.5)$ | $(a_5) = (0.0, 0.0, -0.5)$ | $(a_6) = (-2.0, -2.0, -0.5)$ |
| $(a_7) = (-2.0, -2.0, 0.5)$ | $(a_8) = (-2.0, -2.0, 1.5)$ | $(a_9) = (-2.0, -1.0, -1.5)$ |
| $(a_{10}) = (-2.0, 0.0, -1.5)$ | $(a_{11}) = (-2.0, 0.0, -0.5)$ | $(a_{12}) = (-2.0, 1.0, -1.5)$ |
| $(a_{13}) = (-2.0, -1.0, -0.5)$ | $(a_{14}) = (-2.0, 2.0, -1.5)$ | $(a_{15}) = (-1.0, -2.0, -1.5)$ |
| $(a_{16}) = (-1.0, -2.0, -0.5)$ | $(a_{17}) = (-1.0, -2.0, 1.5)$ | $(a_{18}) = (-1.0, 0.0, -1.5)$ |
| $(a_{19}) = (-1.0, -1.0, -1.5)$ | $(a_{20}) = (-1.0, 2.0, -1.5)$ | $(a_{21}) = (0.0, -2.0, -1.5)$ |
| $(a_{22}) = (0.0, -2.0, -0.5)$ | $(a_{23}) = (0.0, -2.0, 1.5)$ | $(a_{24}) = (0.0, -1.0, -1.5)$ |
| $(a_{25}) = (0.0, -1.0, -0.5)$ | $(a_{26}) = (0.0, 1.0, -1.5)$ | $(a_{27}) = (-1.0, -2.0, -1.5)$ |
| $(a_{28}) = (-1.0, -2.0, -0.5)$ | $(a_{29}) = (1.0, 0.0, -1.5)$ | $(a_{30}) = (-1.0, 0.0, -1.5)$ |
| $(a_{31}) = (0.0, -2.0, -1.5)$ | $(a_{32}) = (0.0, -2.0, 1.5)$ | $(a_{33}) = (-2.0, -1.0, -0.5)$ |

*Table 4: Different phases with the labels corresponding to each configuration, in Figs.5 (a, b, c, d) and Figs.6 (a, b, c, d, e, f).*

In Fig.5a, we present in the plane (H, D) the stable configurations for a specific value of different exchange coupling interactions: $J_{Co-Co}=J_{Fe-Fe}=J_{Cr-Cr}=J_{Co-Fe}=J_{Co-Cr}=J_{Fe-Cr}=1$. The total number of possible stable configurations is $(2 \times S+1) \times (2 \times \sigma+1) \times (2 \times Q+1)=100$, with $S=2$, $\sigma=2$ and $Q=3/2$. From these 100 stable configuration only the configurations: $(a_1)$, $(a_2)$, $(a_3)$ and $(a_4)$ and their opposites $(-a_1)$, $(-a_2)$, $(-a_3)$ and $(-a_4)$ are found to be stable in this figure, see table 4. A perfect symmetry is present in this figure according to the axis H=0.

On the other hand, we have plotted in Fig.5b corresponding to the plane (D, $J_{Fe-Fe}$) the obtained results for H=1 and $J_{Co-Co}=J_{Cr-Cr}=J_{Co-Fe}=J_{Co-Cr}=J_{Fe-Cr}=1$. This figure represents 4 regions. The region 1 where the phases $(a_1)$, $(a_3)$, $(a_{15})$ and $(a_{22})$ are found to be stable. While, the region 2 corresponds to the stable configurations: $(a_1)$, $(a_3)$, $(a_9)$, $(a_{10})$, $(a_{18})$ and $(a_{26})$. The configurations $(a_1)$, $(a_9)$, $(a_{12})$, $(a_{19})$ and $(a_{26})$ are present in region 3. The fourth region contains the stable configurations: $(a_3)$, $(a_{22})$, $(a_{25})$ and $(a_{26})$.

In order to inspect the stable configurations in the plane (D, $J_{Cr-Cr}$), we present in Fig.5c the obtained stable configurations for H=1 and $J_{Co-Co}=J_{Fe-Fe}=J_{Co-Fe}=J_{Co-Cr}=J_{Fe-Cr}=1$. This figure presents also 4 regions. The region 1 contains the stable phases: $(a_1)$, $(a_2)$, $(a_3)$, $(a_4)$, $(a_5)$ and $(a_{30})$. In the region 2, we found the stable configurations $(a_1)$, $(a_2)$, $(a_5)$ and $(a_6)$. The third region contains only three stable configurations, namely: $(a_4)$, $(a_5)$ and $(a_6)$. In the last region, we found the stable configurations: $(a_2)$, $(a_3)$, $(a_4)$, $(a_5)$ and $(a_{30})$.

Similarly to Figs.5 (a, b, c), we plot in Figs.6 (a, b, c, d, e, f) the stable configuration in different planes corresponding to different physical parameters. In fact, we illustrate in Fig.6a the obtained ground state phase diagram of the studied system, in the plane ($J_{Co-Co}$, $J_{Fe-Fe}$) for $J_{Cr-Cr}=J_{Co-Fe}=J_{Co-Cr}=J_{Fe-Cr}=1$ and H=D=1. This figure shows 4 regions: Region 1, containing only the two stable phases $(a_{21})$ and $(a_{27})$. Region 2, showing the 9 stable configurations, namely: $(a_1)$, $(a_4)$, $(a_{10})$, $(a_{12})$, $(a_{15})$, $(a_{21})$, $(a_{26})$, $(a_{30})$ and $(a_{33})$. In region 3, we found the three stable configurations: $(a_{12})$, $(a_{26})$ and $(a_{30})$. In the last region, we found four stable configurations: $(a_{21})$, $(a_{26})$, $(a_{27})$ and $(a_{30})$.

In the plane ($J_{Co-Co}$, $J_{Co-Fe}$) corresponding to Fig.6b, we supply the stable configurations for the specific values of the exchange coupling interactions for $J_{Fe-Fe}=J_{Cr-Cr}=J_{Co-Cr}=J_{Fe-Cr}=1$ and H=D=1. From this figure it is found that the eight stable configurations: $(a_{15})$, $(a_{31})$, $(a_{21})$, $(a_{27})$, $(a_{32})$, $(a_{20})$, $(a_{14})$ and $(a_{32})$ are present for $J_{Co-Co}$ taking negative values. When exploring the stable phases in the plane ($J_{Co-Co}$, $J_{Co-Cr}$), we provide in Fig.6c, the obtained results for $J_{Fe-Fe}=J_{Cr-Cr}=J_{Co-Fe}=J_{Fe-Cr}=1$ and H=D=1. For $J_{Co-Co}$ taking negative values, the stable configurations: $(a_1)$, $(a_{15})$, $(a_{21})$, $(a_{27})$, $(a_{17})$, $(a_8)$ and $(a_{23})$ are all present in this region. While, for $J_{Co-Co}>0$, the only stable phases are: $(a_1)$, $(a_8)$, $(a_{13})$, $(a_{21})$. In order to inspect the effect $J_{Fe-Fe}$, we plot Fig.6d in the plane ($J_{Fe-Fe}$, $J_{Cr-Cr}$) the obtained stable configurations for $J_{Co-Co}=J_{Co-Fe}=J_{Co-Cr}=J_{Fe-Cr}=1$ and H=D=1. The only stable phases: $(a_{10})$, $(a_{12})$, $(a_1)$, $(a_6)$ and $(a_9)$ are engaging the four different areas. When combining the effect the two exchange coupling interactions $J_{Cr-Cr}$ and $J_{Co-Co}$ on the stable configurations, we illustrate in Fig.6e, the obtained results for $J_{Fe-Fe} = J_{Co-Fe} = J_{Co-Cr} = J_{Fe-Cr} = 1$. This figure expresses the stable phases: $(a_1)$, $(a_6)$, $(a_{10})$, $(a_{15})$, $(a_{16})$, $(a_{21})$, $(a_{22})$, $(a_{27})$ and $(a_{28})$ in four different areas, see Fig.6e.

The consequence of the varying the exchange coupling interaction $J_{Fe-Cr}$ on the stable phases is illustrated in Fig.6f, plotted in the plane $(J_{Fe-Cr}, J_{Cr-Cr})$ for $J_{co-Co} = J_{Fe-Fe} = J_{Co-Fe} = J_{Co-Cr} = 1$ and $H=D=1$. Only four stable configuration are found in this figure, namely: $(a_7)$, $(a_8)$, $(a_1)$ and $(a_6)$. These phases are found to be stable in four different regions, see Fig.6f.

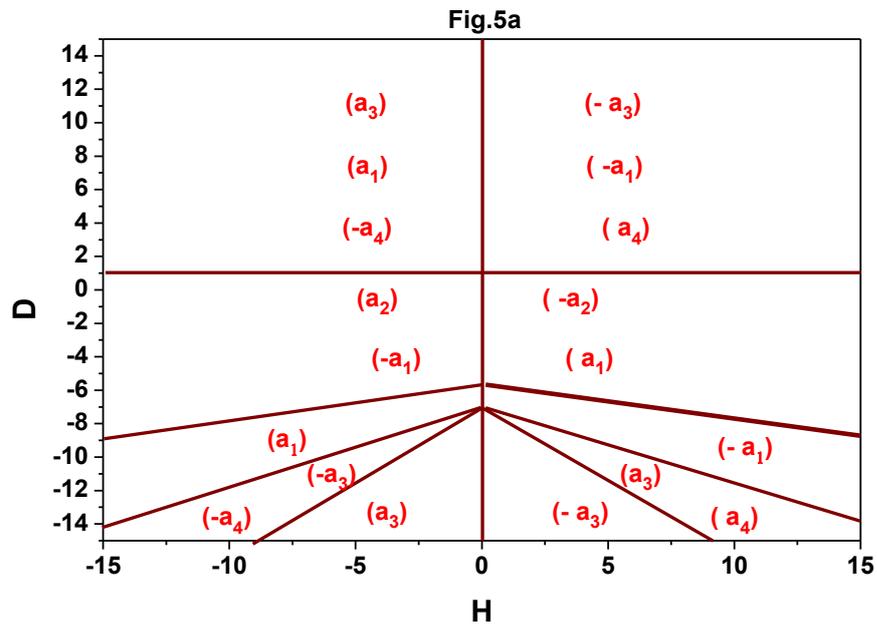

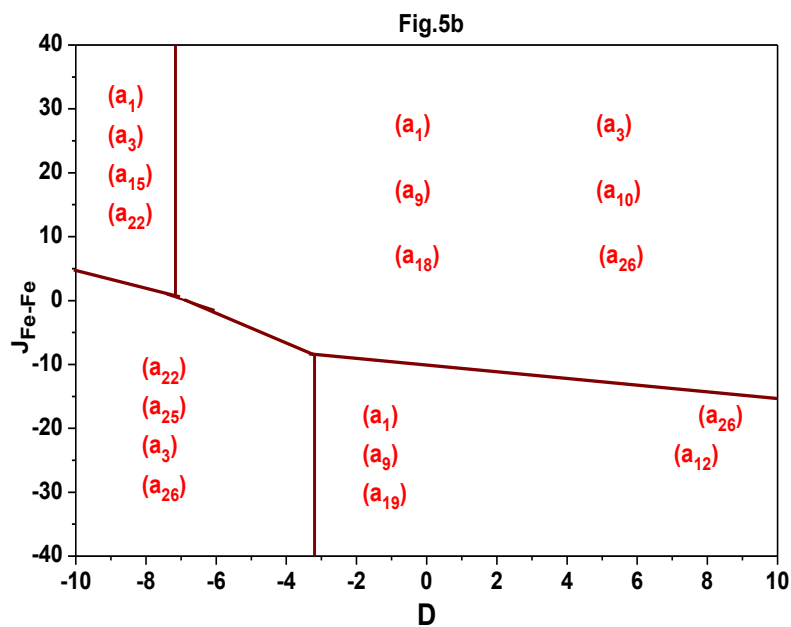

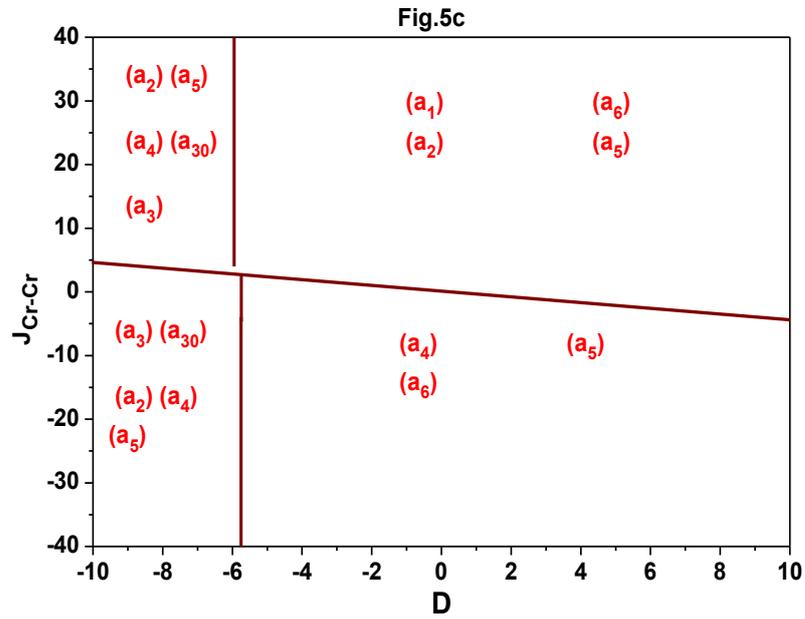

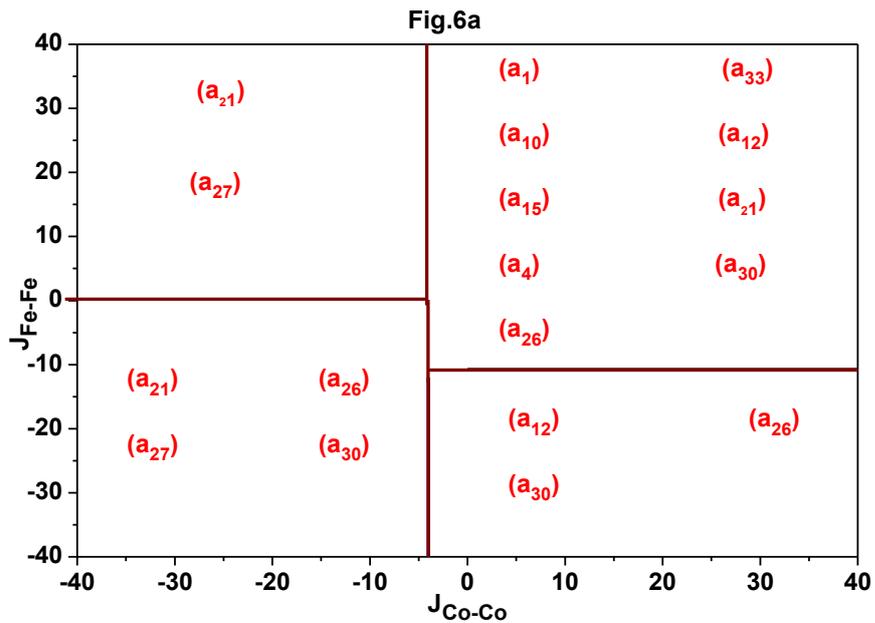

*Fig.5: Ground state phase diagrams of the studied system:*

*(a) in the plane (H, D) for $J_{Co-Co}=J_{Fe-Fe}=J_{Cr-Cr}=J_{Co-Fe}=J_{Co-Cr}=J_{Fe-Cr}=1$;*

*(b) in the plane (D, $J_{Fe-Fe}$) for H=1 and $J_{Co-Co}=J_{Cr-Cr}=J_{Co-Fe}=J_{Co-Cr}=J_{Fe-Cr}=1$;*

*(c) in the plane (D, $J_{Cr-Cr}$) for H=1 and $J_{Co-Co}=J_{Fe-Fe}=J_{Co-Fe}=J_{Co-Cr}=J_{Fe-Cr}=1$;*

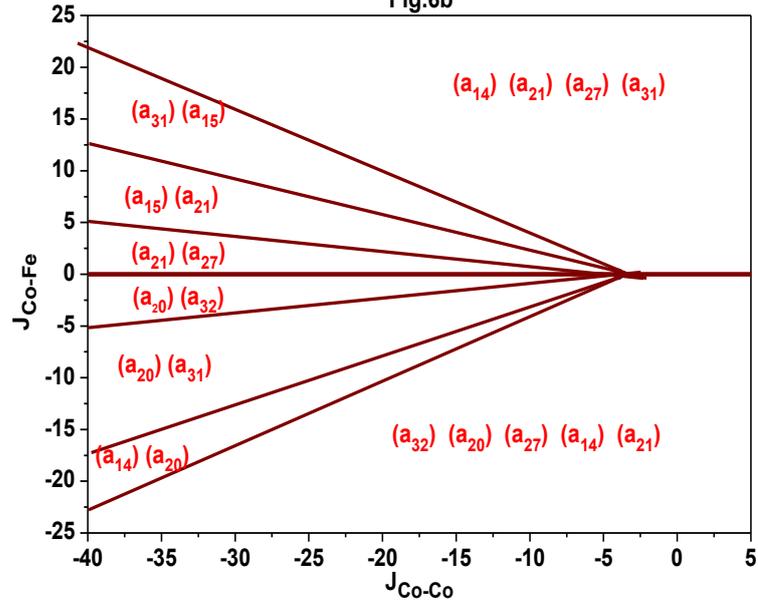

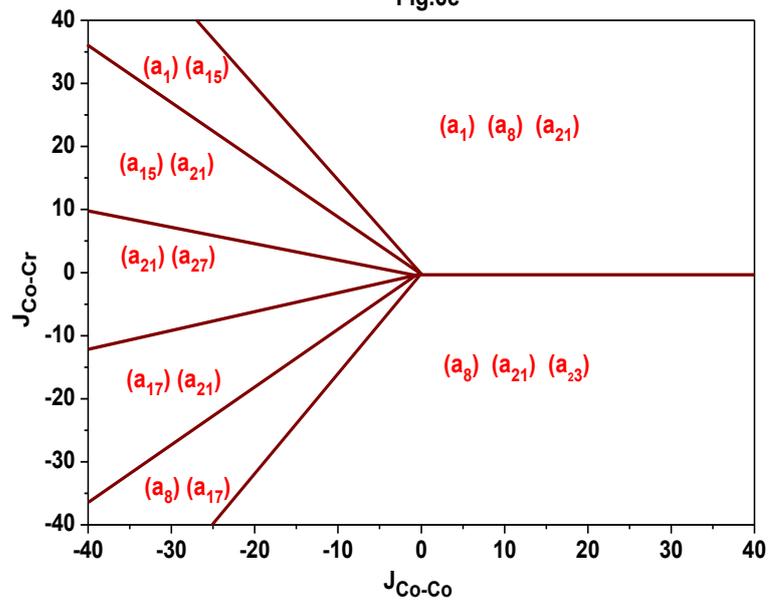

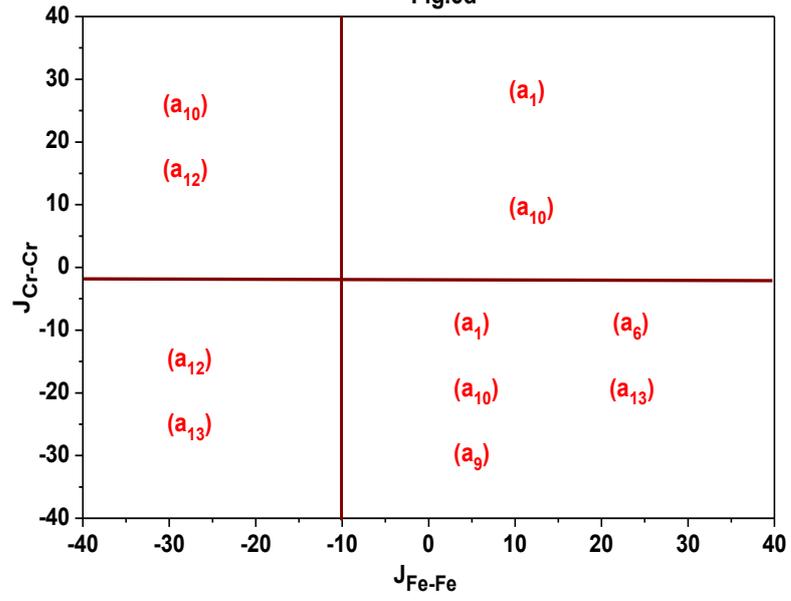

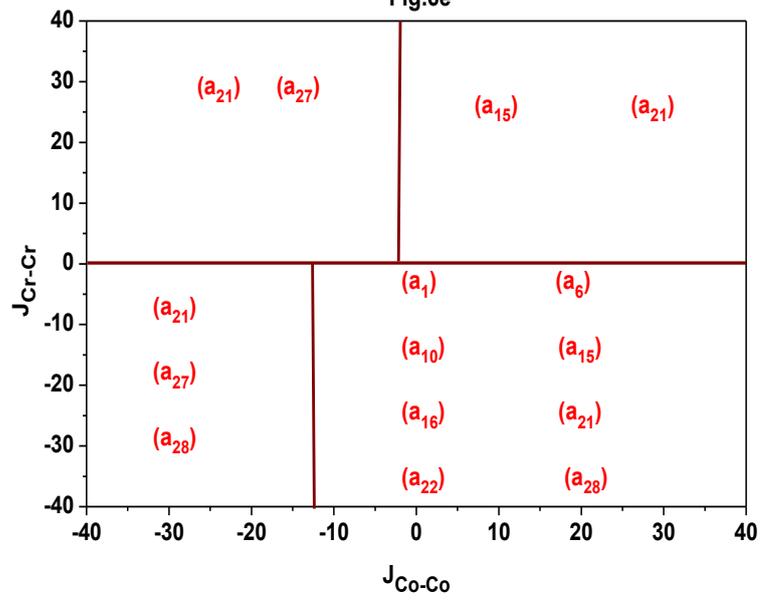

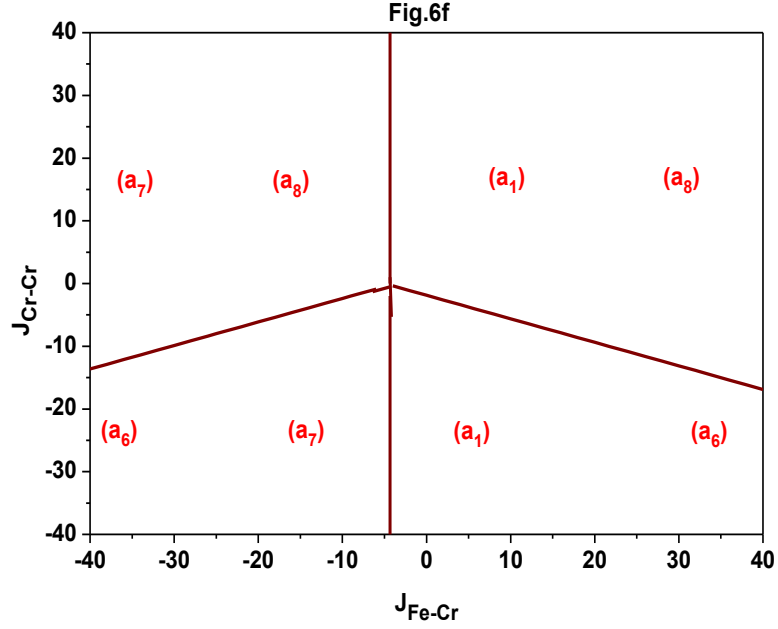

*Fig.6: Ground state phase diagrams of the studied system for H=D=1:*

*(a) in the plane ($J_{Co-Co}$, $J_{Fe-Fe}$) for $J_{Cr-Cr}=J_{Co-Fe}=J_{Co-Cr}=J_{Fe-Cr}=1$;*

*(b) in the plane ($J_{Co-Co}$, $J_{Co-Fe}$) for $J_{Fe-Fe}=J_{Cr-Cr}=J_{Co-Cr}=J_{Fe-Cr}=1$;*

*(c) in the plane ($J_{Co-Co}$, $J_{Co-Cr}$) for $J_{Fe-Fe}=J_{Cr-Cr}=J_{Co-Fe}=J_{Fe-Cr}=1$;*

*(d) in the plane ($J_{Fe-Fe}$, $J_{Cr-Cr}$) for $J_{Co-Co}=J_{Co-Fe}=J_{Co-Cr}=J_{Fe-Cr}=1$;*

*(e) in the plane ($J_{Co-Co}$, $J_{Cr-Cr}$) for $J_{Fe-Fe}=J_{Co-Fe}=J_{Co-Cr}=J_{Fe-Cr}=1$;*

*(f) in the plane ($J_{Fe-Cr}$, $J_{Cr-Cr}$) for $J_{co-Co}=J_{Fe-Fe}=J_{Co-Fe}=J_{Co-Cr}=1$;*

### b. Monte Carlo study of the equiatomic quaternary Heusler alloy CoFeCrP

For each spin configuration, we use $10^5$ Monte Carlo steps (MCS) in order to gain the equilibrium the system. At every one MCS, all the sites of the system are swept and a single-spin flip attempt is made. The flips are accepted or rejected dependent on the Boltzmann insights, under the Metropolis calculation. At equilibrium, averages of the partial and total magnetizations, the partial and total magnetic susceptibilities, the energy of the system and the specific heat have been counted. For each iteration, we estimate the internal energy per site given by the Eq:

$$E_{tot} = \frac{1}{N} < \mathcal{H} > \quad (4)$$

Where, $N=N_S+ N_\sigma+ N_Q$, with $N_S$, $N_\sigma$ and $N_Q$ being the number of magnetic atoms S, σ and Q, respectively.

The total magnetizations is :

$$m_T = \frac{N_S m_S + N_\sigma m_\sigma + N_Q m_Q}{N_S + N_\sigma + N_Q} \quad (5)$$

The partial magnetizations are calculated by:

$$m_S = \langle \frac{1}{N_S} \sum_i S_i \rangle \quad (6)$$

$$m_\sigma = \langle \frac{1}{N_\sigma} \sum_i \sigma_i \rangle \quad (7)$$

$$m_Q = \langle \frac{1}{M_Q} \sum_i Q_i \rangle \quad (8)$$

The total and partial susceptibilities:

$$\chi_S = \frac{\langle m_S^2 \rangle - \langle m_S \rangle^2}{K_B T} \quad (9)$$

$$\chi_\sigma = \frac{\langle m_\sigma^2 \rangle - \langle m_\sigma \rangle^2}{K_B T} \quad (10)$$

$$\chi_Q = \frac{\langle m_Q^2 \rangle - \langle m_Q \rangle^2}{K_B T} \quad (11)$$

$$\chi_T = \frac{\langle m_T^2 \rangle - \langle m_T \rangle^2}{k_B T} \quad (12)$$

The total specific heat:

$$Cv = \frac{\langle E_T^2 \rangle - \langle E_T \rangle^2}{k_B T^2} \quad (13)$$

Where, $k_B$ is the Boltzmann constant ($k_B=1$), T being the absolute temperature.

In Fig.7a, we illustrate the behavior of the total magnetization of the equiatomic quaternary Heusler alloy CoFeCrP. From this figure, it is found that for very low temperature values, we gained the ground state value: (S+σ+Q)/3= (2+2+3/2)/3=1.83. This is a good agreement with the ground state phase diagrams presented in Figs. 5a, 5b and 5c. This figure shows also the fact that when increasing the unified exchange coupling interaction reduced crystal field d, while d=D/J, the saturation of the magnetizations are delayed. In connection with Fig.7a, we plot in Fig.7b the corresponding susceptibilities as a function the temperature for the same values reduced crystal field d. This figure supports the fact that the transition temperature is increased towards higher values of the temperature when increasing the parameter J. This is well reflected by displacement of the peak of the susceptibilities as it is shown in Fig.7b. Table 5 summarizes the variation of the transition temperature when increasing the exchange coupling values from T=200 K to T=750 K, when d increases from d=0.1 to d=0.4, respectively. The last values of T is in good accord with critical temperature of the equiatomic quaternary Heusler alloy CoFeCrP in the literature, $T_c$ =759 K in the Ref. [36] and $T_c$ =635 K in the Ref. [50].

To complete this survey, we provide in Fig.7c the behavior of the specific heat as a mapping of temperature for the same values of the reduced crystal field d as in Figs.7a and 7b. In fact, this figure confirms also the increasing of the transition temperature when increasing the exchange coupling interaction J. In fact, the peak of the specific heat increases with the temperature as it is well illustrated in Fig.7c. Finally, Fig.8 presents the behavior the total energy as a function of the temperature for H=0, D=1 and d=0.1. As expected, the energy reaches its saturation at high temperature values.

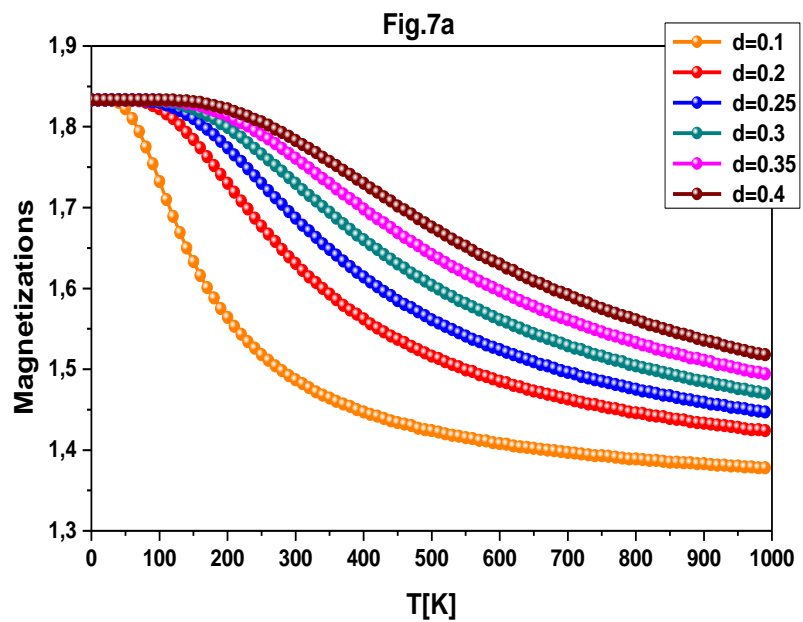

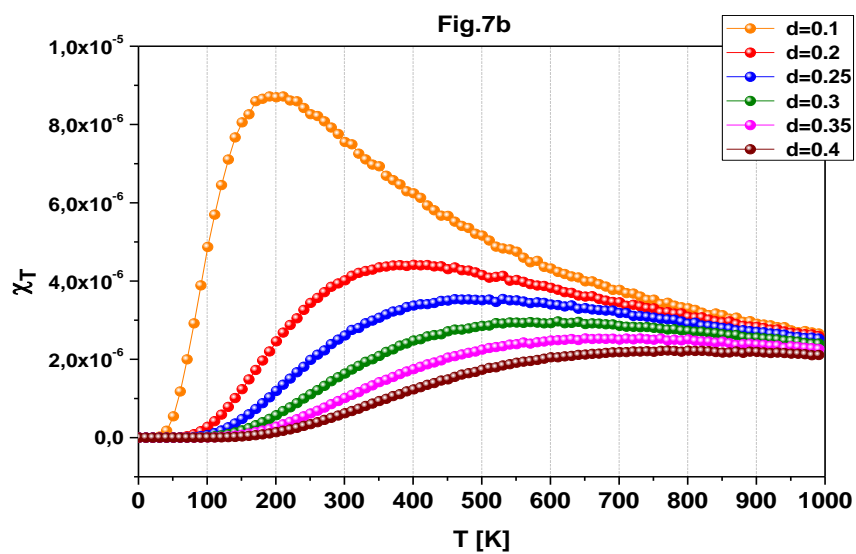

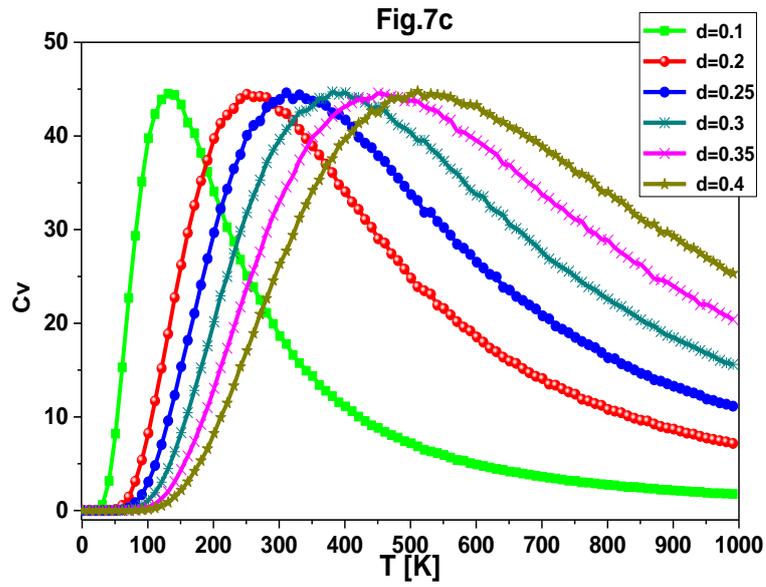

*Fig.7: Thermal behavior for H=0 and **Reduced crystal field** d; of the total magnetization (a), the total susceptibility (b), total specific heat (c).*

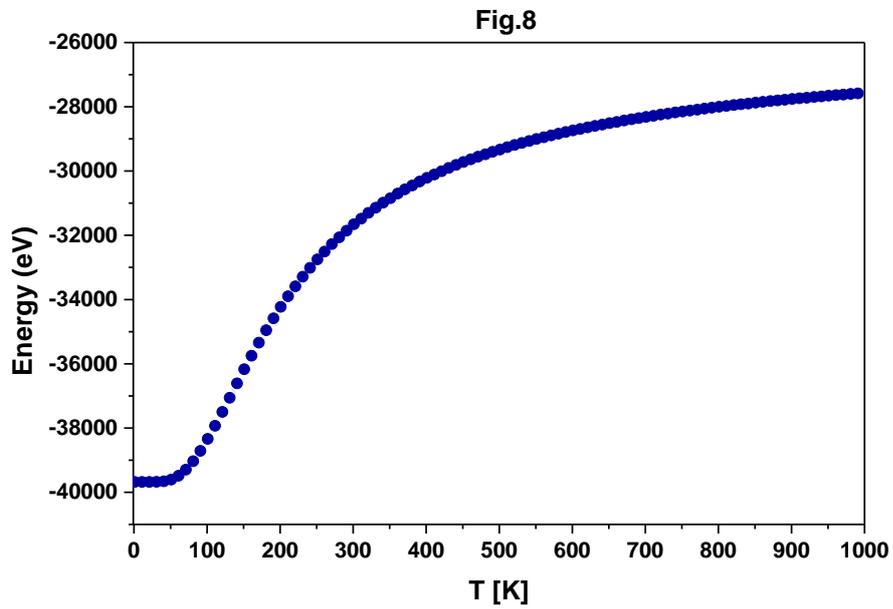

*Fig.8: Total energy of the equiatomic quaternary Heusler alloy CoFeCrP for H=0 and reduced crystal field value d=0.1.*

## IV. Conclusion

To conclude, we used the first principles calculations using the spin polarized density functional theory with the Pseudo-Potential Plane Wave method PP-PW to predict the structural, electronic and magnetic properties of the new equiatomic quaternary Heusler alloy CoFeCrP. The stable structure is the type (I). The band structures and the state density confirm the half-metallic ferromagnetic behavior of this compound CoFeCrP.

We show in different planes, the physical parameters corresponding to the stable configurations for specific values of different exchange coupling interactions. The total number of possible stable configurations is $(2xS + 1) \times (2x\sigma + 1) \times (2xQ + 1) = 100$, with the spin moment values: $S = 2$, $\sigma = 2$ and $Q = 3/2$. A perfect symmetry is present in the figure drawn in the plane (H, D) along the axis $H = 0$. The behavior of the total magnetization of the equiatomic quaternary Heusler alloy CoFeCrP is studied as a function of temperature, using Monte Carlo simulations.

The corresponding susceptibilities as a function of temperature for the same values of the exchange coupling interactions confirm the fact that the transition temperature is increased towards higher values of the temperature during the increase of the parameter J. The value of the transition temperature obtained, for $d = 0.4$ is in good agreement with that given by ref. [36, 46] To complete this study, we provide the dependence of specific heat and total energy, as a function of temperature. The calculated peak of specific heat increases with increasing temperature values. As expected, energy reaches saturation at high temperature values.